\documentclass[aps,prb,twocolumn,groupedaddress,showpacs]{revtex4-1}
\usepackage{graphicx}
\usepackage{amsmath,bm}

\begin{document}

\title{Multiband Electronic Structure of Magnetic Quantum Dots: Numerical Studies}

\author{D. Rederth, R. Oszwa\l dowski, and A. G. Petukhov}
\affiliation{South Dakota School of Mines and Technology, Rapid City, USA\\Email: daniel.rederth@mines.sdsmt.edu}
\author{J.~M.~Pientka}
\affiliation{St.~Bonaventure University, New York, USA}

\date{\today}

\begin{abstract}
Semiconductor quantum dots (QDs) doped with magnetic impurities have been a focus of continuous research for a couple of decades. A significant effort has been devoted to studies of magnetic polarons (MP) in these nanostructures. These collective states arise through exchange interaction between a carrier confined in a QD and localized spins of the magnetic impurities (typically: Mn). We discuss our theoretical description of various MP properties in self-assembled QDs. We present a self-consistent, temperature-dependent approach to MPs formed by a valence band hole. We use the Luttinger-Kohn k$\cdot$p Hamiltonian to account for the important effects of spin-orbit interaction. 
\end{abstract}


\maketitle

\section{Introduction\label{}}
With spintronics and quantum computing as the driving forces, one of the primary foci of nanomagnetism
and semiconductor spintronics is design and fabrication of magnetic Quantum Dots (QDs) with customized properties.
These nanostructures, based on Dilute Magnetic Semiconductors (DMS), 
are also interesting from a fundamental physics point of view; their description requires a combination of 
quantum and statistical approaches to small systems.

\begin{figure}[h]
\includegraphics[bb= 10 80 550 240,clip,scale=0.55]{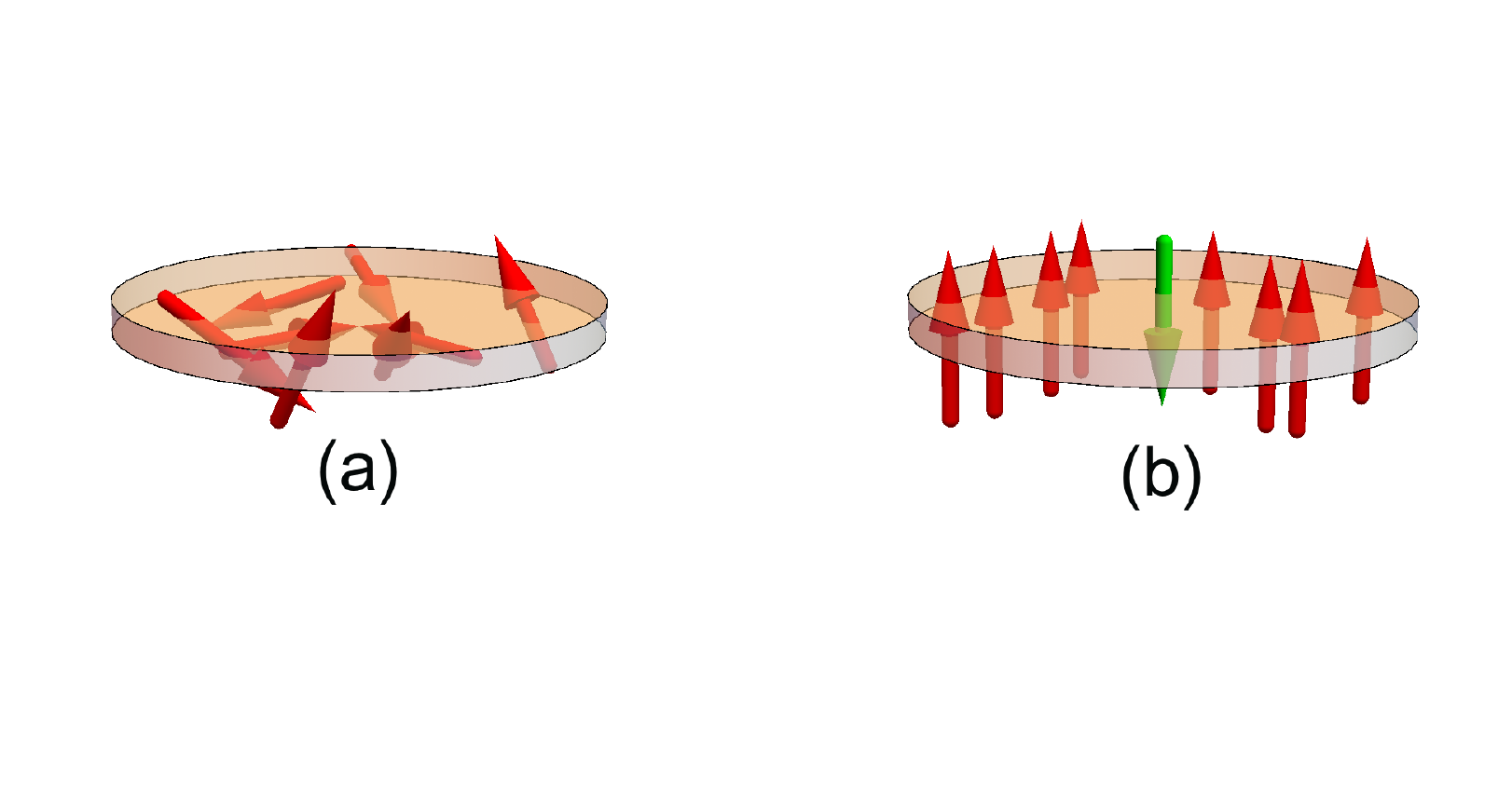}
\caption[Dilute Magnetic Semiconductor Quantum Dot]{Dilute magnetic semiconductor quantum dot. 
(a) Situation without a hole: the Mn ions' spins point in random directions.
(b) Situation when a hole is confined in the QD: the Mn ions' spins align anti-parallel to the hole spin forming a magnetic polaron.}\label{MnHole}
\end{figure}

The magnetic properties of DMS are introduced by the transition-metal ions (such as manganese). \cite{Furdyna1988:JAP} In bulk samples of DMS, alignment of Mn spins is typically achieved through an \textit{external} magnetic field. 
An alternative scenario may be realized in magnetic QDs charged with carriers. Owing to their strong confinement, the exchange interaction of these carriers with Mn ions is enhanced. This obviates the need of external magnetic field,  replaced by spin-density of the trapped carriers.
This effective \textit{internal} field may be large enough to strongly align the Mn spins,\cite{Rice2016:NN}
resulting in a magnetized quantum dot, Fig.~\ref{MnHole}. 
At the same time, the ground-state energy of the carrier is lowered,
and spin degeneracy of energy levels is lifted.
This combination of effects is referred to as formation of a magnetic polaron (MP). 

Evidence of MP formation is provided by numerous optical experiments on magnetic II-VI QDs embedded in bulk semiconductor, e.g.~the early reports in  Refs.~\onlinecite{Maksimov2000:PRB,Mackowski2005}.
The first time-resolved studies of this effect in self-assembled DMS QDs were presented in Refs.~\onlinecite{Kuroda2000:JCG,Seufert2002}.   
Later, formation of magnetic polarons was revealed in colloidal magnetic QDs. \cite{Gamelin2009,Rice2016:NN}

In this work, we present a self-consistent, multiband, temperature dependent description of 
hole magnetic polarons. We focus on MPs formed by exchange interaction of a single hole with multiple Mn spins embedded in II-VI QDs, as in the experimental studies Refs.~\onlinecite{Kuo2006:APL,Sellers2010:PRB}. We use the Luttinger-Kohn Hamiltonian to describe quantum states of the hole. For temperature dependence, we introduce a well-controlled mean-field approach. This combination of quantum and statistical descriptions allows us to formulate a self-consistency condition, which reflects the mutual influence of the confined carrier and localized, paramagnetic Mn spins.
 
Magnetic polarons are typically formed at cryogenic temperatures.
The corresponding energy gain is destroyed by 
thermal fluctuations at higher temperatures. 
We present results showing that our method correctly describes this temperature dependence.
Our approach, based on the envelope-function approximation, goes beyond the often employed ``muffin-tin" ansatz for a confined carrier wavefunction. Thus, we are able to reveal an interesting effect: localization (``shrinking") of the hole wavefunction 
in QDs with the above Mn placement.

We note that MP formation is known to occur also in doped bulk DMS systems.  
In that case, a carrier bound to a donor or acceptor  aligns the Mn spins within its effective Bohr radius.\cite{Dietl1983:PRB,Bednarski2012:JPCM}
This scenario, called bound magnetic polaron, has similarity to MPs formed in QDs.  
However, the degree of freedom offered by the tunability of QD confinement is absent in that scenario.

\section{Non-magnetic Hamiltonian}

We employ the Luttinger-Kohn Hamiltonian to describe the hole states:
\begin{equation}
\hat H_{\rm LK} = \left( {\begin{array}{*{20}{c}}
{\hat  P + \hat  Q}&{ - \hat  S}&{\hat  R}&0 \\
{ - \hat  S^*}&{\hat  P - \hat  Q}&0&{\hat  R} \\
{\hat  R^*}&0&{\hat  P - \hat  Q}&{\hat  S} \\
0&{\hat  R^*}&{\hat  S^*}&{\hat  P + \hat  Q} \\
\end{array}} \right),\label{LK4}
\end{equation} 
where all the quantities (containing Luttinger parameters $\gamma_{1,2,3}$) and the phase convention are defined in Ref.~\onlinecite{Chuang}, Cartesian components of wave vector are replaced by partial derivatives.\cite{Bastard}

We take advantage of the Envelope Function Approximation by adding a confining potential, resulting from the band offset at semiconductor heterojunctions. 
We model the  potential by $V({\bf r})I_4$, where $I_4$ is the $4\times 4$ identity matrix. Thus, $V({\bf r})$ is the same for heavy- and light-holes.
It consists of the infinite-well potential in the growth direction, $z$, and in-plane parabolic potential, $\frac{1}{2}m^*\omega^2(x^2+y^2)$, where $m^*=\frac{m_0}{\gamma_1+\gamma_2}$ is the in-plane heavy-hole effective mass.
Altogether, the non-magnetic part of the Hamiltonian is 
$\hat H_0=\hat{H}_{\rm LK}+V({\bf r})I_4.\label{SE}$
\section{Self-consistent Exchange Hamiltonian}
The key element of this work  is the non-linear, temperature dependent Schr{\" o}dinger equation for the effective carrier wavefunction corresponding to the most probable spin fluctuation~\cite{Oszwaldowski2012:PRB}. This equation can be justified as follows. First, we will consider the Hamiltonian describing the 
contact exchange interaction between the fermions and the magnetic ions. It is convenient to express this Hamiltonian in the second-quantized form:
\begin{equation}
\label{Eq:Ising-Ham}
\hat{H}=\hat{H}_0+(\beta/3)\sum_j \hat{S}_{jz}\left(\underline\psi^\dagger_{j}{\hat J}_z\underline\psi_{j}\right)
\end{equation}
where $\underline\psi_j^\dagger=\left(\psi_{j,3/2}^\dagger,\psi_{j,1/2}^\dagger,\psi_{j,-1/2}^\dagger,\psi_{j,-3/2}^\dagger\right)$ is the 
four-component spinor field of a hole with spin $J=3/2$, the spin indexes $\sigma=\pm 3/2$ and $\sigma=\pm 1/2$ correspond to heavy-hole and light-hole states respectively, $\psi^\dagger_{j\sigma}$ and $\psi_{j\sigma}$ are creation and annihilation fermion field operators 
such that $\psi_{j\sigma}\equiv\psi_\sigma(\bf{R}_j)$ where $\bf{R}_j$ is the location of a magnetic ion, 
$\beta$ is the exchange coupling constant, and 
$\hat{S}_{z}$ and $\hat{J}_{z}$ are
the Mn ($S=5/2$)  and the heavy-hole ($J=3/2$) spin operators respectively.  
The Ising Hamiltonian~\eqref{Eq:Ising-Ham} 
is well justified to describe thermodynamic spin fluctuations and formation of the heavy-hole magnetic polarons in quasi two-dimensional quantum dots with strong $g$-factor anisotropy~\cite{Merkulov:1995tl,Vyborny2012}. 

We take advantage of the fact that the Ising Hamiltonian does not contain the double spin-flip processes in which a hole and a Mn impurity exchange a unit spin. It means that the system's wavefunction can be represented as a product of the hole and Mn-spin parts. The resulting hole Hamiltonian will depend on the set of $c$-numbers $S_{jz}$ (spin projections) rather than on the set of the non-commuting spin operators. Second, we assume that the time evolution of the Mn spin subsystem is slow enough to treat it as a static non-uniform exchange field acting upon the hole spins, and consider only the stationary states of the thermalized holes for each configuration of the Mn spin projections. Therefore the partition function of the system can be calculated using Gibbs canonical distribution:
\begin{equation}
\label{Eq.Zexact}
Z=\textrm{Tr}_{S_{jz}}\sum_n\exp\left[\frac{-E_n\left(\{S_{jz}\}\right)}{k_{\rm B}T}\right],
\end{equation}
where $k_{\rm B}$ is the Boltzmann constant, $T$ is temperature, and $n$ is a quantum number labeling the hole eigenvalues that, in turn, depend on $6^N$ $c$-numbers  $S_{jz}$. Thus, in order to calculate the partition function in Eq.~\eqref{Eq.Zexact} one would need to solve $6^N$  replicas of the hole Schr\"odinger equation, which makes the problem intractable.

To overcome this obstacle we will partition the area of the quantum dot with $N$ Mn spins into a set of $N_c$  square blocks (cells), containing few ($N_k$) Mn spins (i.e. $N=N_cN_k$), 
and neglect spatial variation of the hole wavefunction (and the spin density) within each cell. 
For a particular cell $k$ with $N_k$ spins a distribution function of the average dimensionless magnetization, $\mu_k\equiv\bar S_{z}^{(k)}$,
can be expressed as:
\begin{align}
\label{Eq.Y}
Y(\mu_k)&
=\textrm{Tr}_{S_{jz}^{(k)}}\delta\left(\mu_k-N_k^{-1}\sum_j S_{jz}^{(k)}\right)\nonumber\\
&\propto\exp\left[\frac{-G_S(\mu_k/S)}{k_{\rm B}T}\right],
\end{align}
where the Gibbs free energy of $N_k$ non-interacting spins, $G_S(\mu_k/S)$, that can be obtained using Legendre's
transformation, reads:
\begin{equation}
G_S\!(x)\!=\!k_{\rm B}T N_k\!\!\left[xB_S^{-1}\!(x)\!-\!\ln\frac{\sinh\left[(1\!+\!1/2S)B_S^{-1}\!(x)\right]}{\sinh\left[(1/2J)B_S^{-1}\!(x)\right]}\!\right]
\end{equation}
Here $B_S^{-1}(x)$  is the inverse of the Brillouin function  $B_S(x)$.
We note that the distribution function $Y(\mu_k)$  is temperature independent, i.e. purely entropic.

Using the distribution functions $Y(\mu_k)$   we can transform the partition function of Eq.~\eqref{Eq.Zexact} into a multiple integral over continuous variables $\mu_k$:
\begin{equation}
\label{Eq:Zcont}
Z\propto\sum_n\int\prod_{k=1}^{N_c}
d\mu_k
\exp\left[-\frac{\sum_k G_S(\mu_k/S)+E_n(\{\mu_k\})}{k_{\rm B}T}\right]
\end{equation}
The expression in parentheses of Eq.~\eqref{Eq:Zcont} contains two terms: the fermion energy $E_n$ responsible for
the force exerted by the fermions on the magnetic ions and the magnetic term $\sum_k G_S(\mu_k/S)$ responsible for
the restoring  force exerted by the ions on the fermions.  The latter, so-called entropic (or emerging) force has a rather peculiar origin because it is not derived from any physical interaction between the particles, i.e. magnetic ions. Rather it is caused by the tendency of the system to assume the state with the maximum entropy. This is precisely  why 
the magnetic polarons are fundamentally different from the lattice polarons. The entropic forces can be efficiently controlled
by the temperature, magnetic or even electric field. On the contrary, the physical control of the lattice polarons is severely limited. 

For any particular $n$ the integral in Eq.~\eqref{Eq:Zcont}  can be evaluated using the steepest descent method. 
The saddle point equation must be combined with the Hellmann-Feynman theorem  
\[-\frac{\beta\langle \hat{J}_z(k) \rangle}{3} =\frac{\partial E_n(\mu_1,...,\mu_k,...,\mu_{N_c})}{\partial \mu_k},\]
where $\langle \hat{J}_z(k)\rangle=N_k^{-1}\sum_{j=1}^{N_k} \langle \Psi |\psi_j^\dagger\hat{J}_z\psi_j|\Psi\rangle$ is the average hole spin density of the cell.

This leads to the non-linear Schr\"odinger equation~\cite{Oszwaldowski2012:PRB}: \begin{equation}
\frac{\delta \langle \Psi |\hat H_0| \Psi\rangle}{\delta \Psi}+\frac{\beta S }{3}
\sum_k N_k
B_S\left[\frac{S\beta\langle\hat{J}_z{(k)}\rangle}{3k_\textrm{B}T}\right]\frac{\delta\langle\hat{J}_z{(k)}\rangle}{\delta \Psi}
=0,
\label{eq:NLSE}
\end{equation}
Here $|\Psi\rangle$ is the state vector corresponding to the most probable spin fluctuation. The first term in Eq.~\eqref{eq:NLSE} is a variational form of the standard non-magnetic Schr\"odinger equation while the second term describes a non-linear and temperature-dependent contribution of the spin fluctuations induced by the paramagnetic ions.

Analysis of the single magnetic polaron Hamiltonian \cite{Dietl1983:PRB,Wolff:1988uo} shows that the ordered solution of Eq.~\eqref{eq:NLSE} exists at any temperature. 
It means that the exponent in the integrand of Eq.~\eqref{Eq:Zcont} may be expanded around the saddle point and the multiple Gaussian integration with respect to all $\mu_k$ can be carried over. 
Remarkably, the result of this integration coincides with the exact integral calculated by Wolff~\cite{Wolff:1988uo}.
Moreover, one can generalize this procedure to the case of the Heisenberg exchange. 
The Gaussian integration in this case is more intricate because the Hamiltonian possesses continuous rotational symmetry and the expansion of the exponent in Eq.~\eqref{Eq:Zcont} in the vicinity of the saddle point contains two zero-frequency transverse mode in accordance with the Goldstone theorem~\cite{Goldstone:1961vk,Goldstone:1962ty}. 
The proper elimination of the  Goldstone modes allows to complete the Gaussian integration, and the result again agrees with those of 
Wolff~\cite{Wolff:1988uo} and Dietl and Spa\l{}ek~\cite{Dietl1983:PRB}. 
Thus the free energy of a magnetic polaron calculated by means of the steepest descent integration replicates the exact results with no signature of spurious phase transition.
This is because the non-linear Schr{\" o}dinger equation \eqref{eq:NLSE} contains quantum mechanical rather than thermal average of the fermion spin density, contrary to the conventional method previously used in many works on magnetic quantum dots.\cite{Abolfath:2007vu}
The latter approach imposes artificial thermodynamic limit on a nanoscale system leading to spurious results. 
In this case, an attempt to calculate the partition function using the steepest descent method would lead to a divergence $(1-T/T_c)^{-1/2}$ in the vicinity of the critical temperature. 

The coarse-grained variables of Eq.~(\ref{eq:NLSE}) can be replaced with continuous variables in a standard way. 
Thus, the self-consistent approach replaces the Mn spins, $\hat{S}_z$, with their thermal average, i.e.~with magnetization, $m({\bm r}) = SB_S[S\beta\rho_s(\bm r)/3k_{\rm B}T]$, where the spin density $\rho_s(\bm r)=\langle J_z(\bm r)\rangle$ and we replaced coarse-grained cell index $k$ with a continuous variable $\bm r$. 
The exchange Hamiltonian reads~\cite{Oszwaldowski2012:PRB}: 
\begin{equation}
\hat H_{\rm ex}=\frac{1}{3}x_{\rm Mn}\left|N_0\beta\right|m(\bm r)\hat J_z,\label{Hex}
\end{equation}
where $x_{\rm Mn}$ is the Mn (molar) fraction, and $N_0$ is the cation (number) density. (The $x_{\rm Mn}$ values in this paper are effective, i.e., assume no antiferromagnetic coupling between Mn ions.)
The quantities $m(\bm r)$ and $\rho_s(\bm r)$ must be found self-consistently using a continuous version of the nonlinear Schr\"odinger equation~\eqref{eq:NLSE}:
\begin{widetext}
\begin{equation}
\hat H_{0}\underline{F}(\bm r)+\frac{1}{3}x_{\rm Mn}\left|N_0\beta\right|\hat J_zSB_S\left[\frac{S\beta}{3 k_{\rm B} T}\underline{F}^{\dagger}(\bm r)\hat J_z\underline{F}(\bm r)\right]~\underline{F}(\bm r)
=E~\underline{F}(\bm r), \label{NSE}
\end{equation}\end{widetext}
where $\underline{F}(\bm r)=\langle{\bm r}|\Psi\rangle$ is a 4-component spinor:
\begin{equation}
\underline{F}\left({\bm r}\right)=\begin{bmatrix}
F_{3/2}\left({\bm r}\right) \\
F_{1/2}\left({\bm r}\right) \\
F_{-1/2}\left({\bm r}\right) \\
F_{-3/2}\left({\bm r}\right) \\
\end{bmatrix}\label{4spinor}
\end{equation}

\section{Numerical Approach}
We have solved the nonlinear Schr\"odinger Eq.~(\ref{NSE})
with the Finite Difference method. 
The method discretizes the Hamiltonian and envelopes $F$ by dividing the QD into a cubic mesh.\cite{Liu1996,HarrisonFDM}
Since the mesh lengths are on order of the crystal lattice spacing, the derivatives of the envelope functions are well approximated by finite differences. \cite{Liu1996}
This numerical method can be used for any shape of quantum dot, double QDs, and QDs on a wetting layer.\cite{Wojs1996:PRB}

The self-consistent procedure to model the magnetic polaron follows. We start 
with the initial magnetization $m(\bm r)=0$, this models mutual cancellation of random Mn spins.  
This magnetization enters the exchange Hamiltonian, Eq.~\ref{Hex}.
We use the envelopes resulting from 
Eq.~\ref{NSE} to calculate the  
local hole spin density, $\rho^{(i)}_s(\bm r)=\underline{F}^{(i)\dagger}(\bm r) \hat{J}_z \underline{F}^{(i)}(\bm r)$,
which gives a new magnetization [cf Eq.(6) of Ref.~\onlinecite{Oszwaldowski2012:PRB}]:
\begin{equation}\label{selfcon}
m^{(i+1)}\left (\bm r\right)=S B_S\left[\frac{S\beta\rho^{(i)}_s(\bm r)}{3 k_{\rm B} T}\right]
\end{equation}
This form of magnetization is similar to the customary one,\cite{Blundell} except that $\rho_s\left (\bm r\right)$ replaces the external magnetic field.

Eq.~\ref{selfcon} is the self-consistency condition. 
Our goal is to solve  Eq.~\ref{NSE} using a recursive procedure, which loops between the magnetization and spin density.
When the magnetization is within a tolerance  of the previous iteration, then the self-consistency loop ends.  
Our assumption is that after a few iterations, this procedure finds the actual magnetization and a consistent spin 
density.
 
In brief, the self-consistent algorithm \cite{Oszwaldowski2012:PRB} is:
\begin{enumerate}
\item Start with zero position-dependent Mn magnetization, i.e. $m^i=0$, where $i=1$
\item Employing the finite-difference method, solve ${(\hat H_0+\hat H_{\rm ex}(m^i))\underline F^{(i)}=E \underline F^{(i)}}$ with $\hat H_{\rm ex}$ from Eq.~\ref{Hex}
to calculate the next iteration of hole eigenstates (envelope wavefunctions, $F_\sigma$ in Eq.~\ref{4spinor}) 
\label{eqsolve}
\item Calculate the spin density, $\rho^{(i)}_s$, from $\underline{F}^{(i)}$
\item Calculate $m^{(i+1)}$ in Eq.~\ref{selfcon} using $\rho^{(i)}_s$
\item Find the maximum, with respect to position, of the absolute value of 
differences $\left| m^{(i)}-m^{(i+1)}\right|$. If it is more than a specified 
tolerance $\epsilon$, replace $m^{(i)}$ with $m^{(i+1)}$ 
in the subsequent iteration, ($i\rightarrow i+1$),
go to point \ref{eqsolve}
\item 
Otherwise, the iteration loop ends, effectively solving the nonlinear Schr\"odinger equation Eq.~\ref{NSE}
\end{enumerate}
We used $\epsilon=0.5\times10^{-3}$. As a result, we obtain: the energy of the ground state, its envelopes, and the magnetization profile. 
These quantities indicate MP formation for suitable system parameter regimes.
\section{Self-Consistent Results\label{Sec.Results}}  
In this section, we present our numerical results obtained using the above approach.
Our standard system is a Cd$_{97\%}\text{Mn}_{3\%}$Te QD.
The parabolic potential is given by $\hbar\omega=30$ meV, corresponding to the in-plane 
characteristic length, $\xi_0=\sqrt{\frac{\hbar}{m^*\omega}}$= 4.19 nm, where $\hbar$ is the Dirac constant. 
The distance between the infinite barriers, i.e., the QD height, is $h_{\rm QD}=3$~nm.
The Luttinger parameters are $\gamma_1$=5.3, $\gamma_2$=1.62, $\gamma_3$=2.1. \cite{Said1990:pssb}
The exchange coupling constant, $\beta$, is given by $\vert N_0\beta\vert=0.88$ eV. \cite{Furdyna1988:JAP}

The energy gain due to the magnetic polaron, $E_{\rm MP}$, is defined as the difference between the ground-state energy of a non-magnetic and magnetic QD: $E_{\rm MP}=E_{\rm GS}(x_{\rm Mn}=0)-E_{\rm GS}(x_{\rm Mn}>0)$.
This energy gain is realized only at low temperatures.
\begin{figure}[h] 
\centering
\includegraphics[scale=0.63]{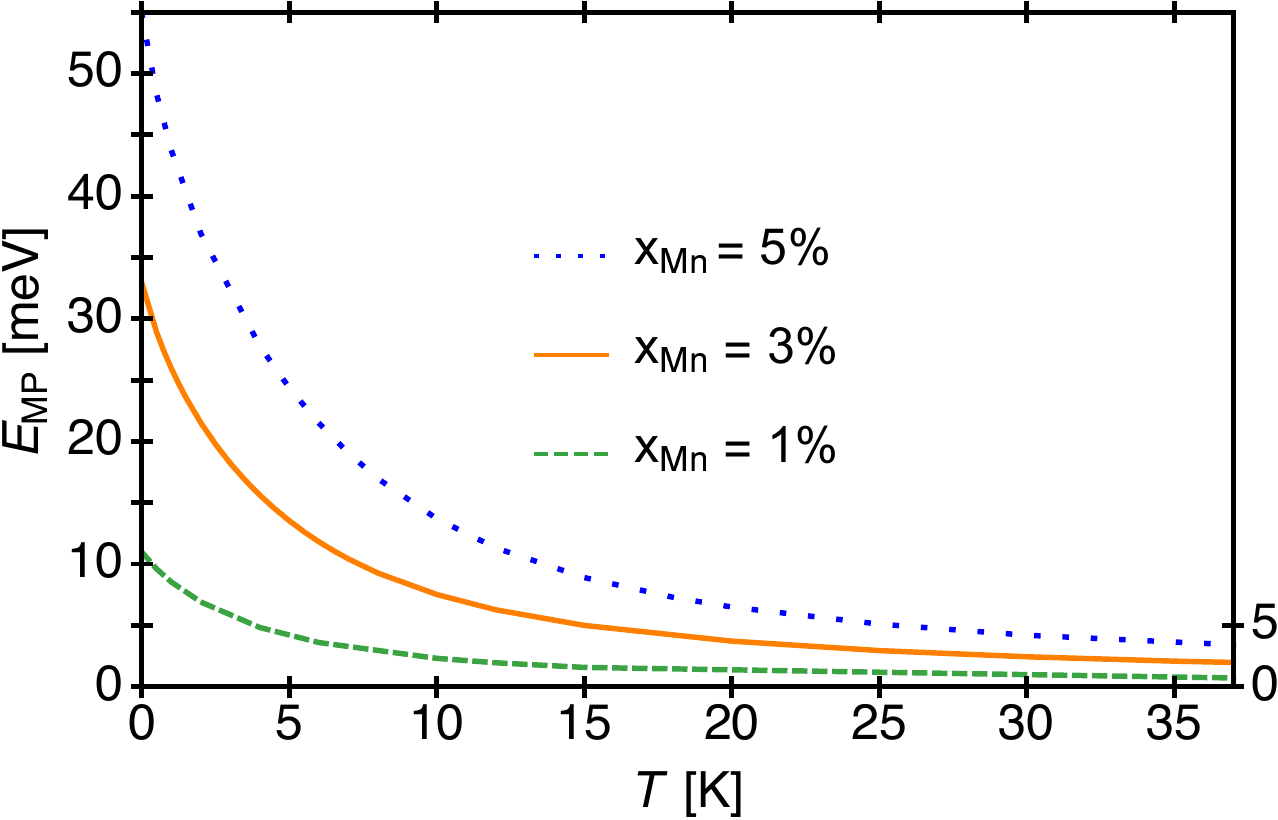}
\caption[The Energy Gain from the Magnetic Polaron, $E_{\rm MP}$]{The energy gain from the magnetic polaron, $E_{\rm MP}$, for different Mn contents.  The MP energy is at maximum at $T=0$ K.  
}
\label{emp2}
\end{figure}
At higher temperatures, thermal excitation overcomes the exchange energy gain and the QD relaxes into a non-magnetic state.
This temperature dependence, as well as the dependence on $x_{\rm Mn}$ is shown in Fig.~\ref{emp2}.

\begin{figure} 
\centering
\includegraphics[scale=0.85]{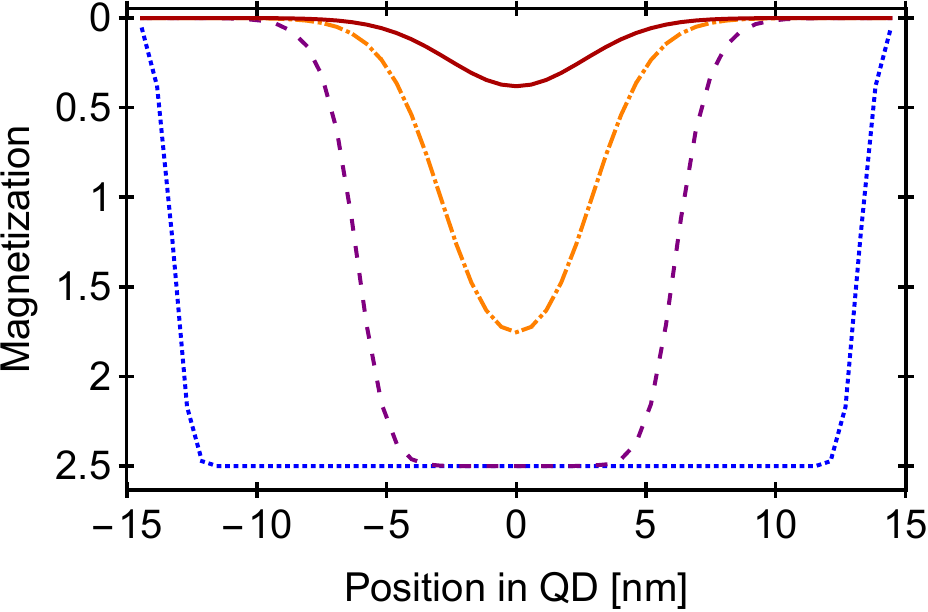}
\caption[Mn Magnetization Profile]{Mn magnetization profile, m($\bm r$), 
vs.~temperature. 
At $T=0$ K (dotted line), the Mn spins are aligned and parallel, causing the magnetization to be saturated. 
At $T=0.5$ K (dashed line), the energy gain from the aligned Mn spins at the QD center causes the magnetization to be saturated only in the QD center, while the tails of the profile undergo thermal disruption.  
At $T=8$ K (dot-dashed line), the magnetization is still evident; at $T=40$ K (solid line), the magnetization is almost lost to temperature.}\label{fig:mvT}
\end{figure}

Fig.~\ref{fig:mvT} shows our results for the position-dependent magnetization $m({\bm r})$.
This quantity has a strong temperature dependence. It saturates at $m=5/2$ for low temperatures. The region with saturated $m$ is centered on the QD center, its volume decreases with increasing temperature.

\begin{figure}[h]
\centering 
\includegraphics[scale=0.666]{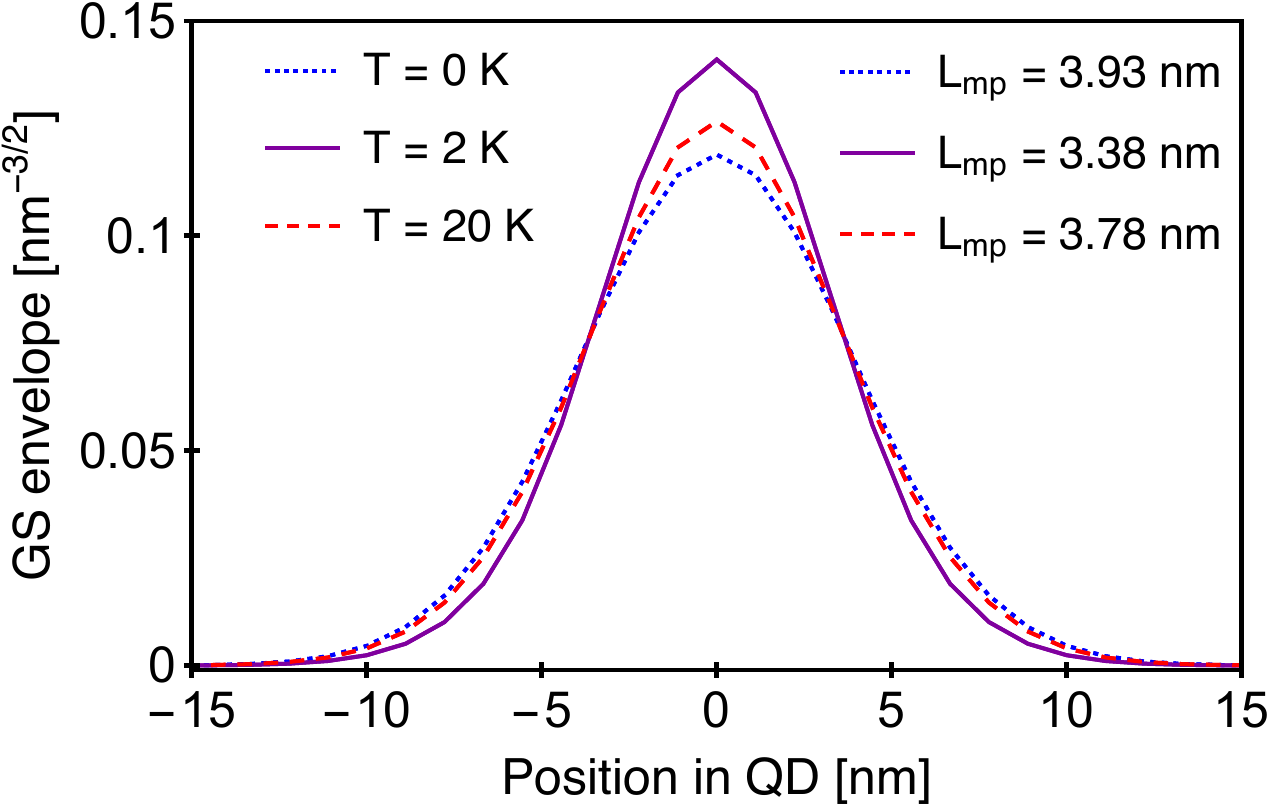}
\caption[Wavefunction Localization]{At $T=0$ (dotted line), the wavefunction envelope does not localize 
and its width is very close to that of the non-magnetized case. 
The localization effect is strong at $T=2$~K (solid line).  At $T=20$~K (dashed line), the envelope starts to relax into a non-magnetic state due to thermal excitation of the system.}
\label{wavePinch}
\end{figure}

Fig.~\ref{wavePinch} demonstrates an interesting effect found in our simulations: ``shrinking" of the envelopes in a temperature range. Because the Mn spins coupled to the tail of the wavefunction (i.e., on the QD periphery) are more prone to thermal excitation with rising temperature, the wavefunction localizes to the center, thus maintaining some of the exchange energy gain through a stronger alignment of Mn spins in the central region.\cite{Pientka2015:PRB}

We quantify the localization effect through the in-plane ``envelope width", $L_{\rm mp}$, defined as
\begin{equation}
L_{\rm mp}=\sqrt{\int \sum_{\sigma}\left|F_\sigma(x,y,z)\right|^2(x^2+y^2)dxdydz}
\label{Lmp}
\end{equation}
For our standard QD, the in-plane envelope localization is 14 times stronger 
than out-of-plane (defined analogously to Eq.\ref{Lmp}).
The out-of-plane envelope localization is about 1\%, thus negligible to our study.   

\begin{figure}[h]
\includegraphics[scale=0.55]{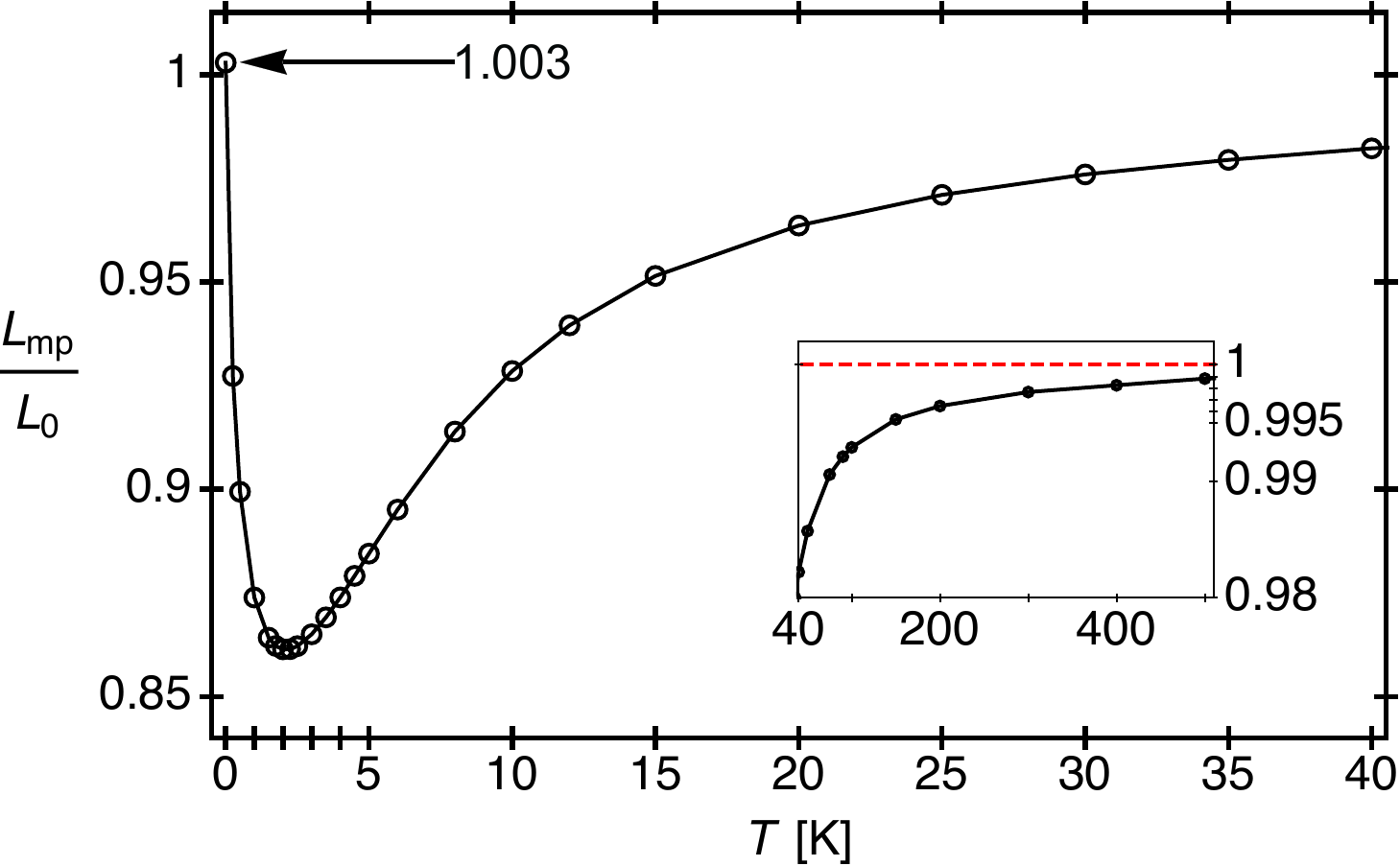} 
\caption{
Localization due to magnetic polaron vs. temperature for a Cd$_{1-x}$Mn$_x$Te QD with $x=3\%$ Mn.
The envelopes exhibit strong localization for temperatures between 1 and 3~K.
Inset: localization decreases at high temperatures, so the ratio becomes 1.}\label{RenormRatio}
\end{figure}

Fig.~\ref{RenormRatio} shows the numerical temperature dependence of $L_{\rm mp}$.
For clarity, we normalize $L_{\rm mp}$ to the non-magnetic width, $L_{\infty}$, which is realized at $T \to \infty$ or, equivalently, for $x_{\rm Mn}=0$.
At $T=0$, the mixing of the light- and heavy-holes leads to a small difference of the envelope with respect to the high $T$ limit: 
$L_{\rm mp}(T=0)/L_{\infty}=1.003$.
(The zero--$T$ and high--$T$ envelopes are exactly equal in the single-band approximation, where the light- and heavy-holes do not mix, so that $L_{\infty}=L_{\rm mp}(T=0)=\xi_0$.)

At high temperatures, the magnetization is thermally destroyed. Since there is no more energy gain from the envelope localization, the system relaxes to a nonmagnetic state, see Fig.~\ref{RenormRatio} inset.

\section{Delocalization: M\MakeLowercase{n} Outside QD\label{Sec.CEPartFunc.}}

We have shown above that Mn placed in QDs produces a localization effect on the carrier wave function.
In the seminal experiment reported by Seufert et al.,\cite{Seufert2002} as well as others,\cite{Barman2015:PRB} the Mn ions were placed (nominally) outside of QDs i.e., in the barrier. What effect will this Mn position have on the carrier wave function confined inside the QD? 

To study this problem, we consider the saturated regime ($T$ = 0 K), where the Mn spins are fully spin polarized.
We use a use a simple single-band model, in which the heavy-hole envelope is the ground-state solution of a 2D harmonic oscillator in the $x-y$~plane, times a normalized sine along the $z-$axis (consistent with the potential assumed in Sec.~\ref{Sec.Results})
 \begin{equation}
F_{\rm HH}(r,z) = \frac{1}{\sqrt{\pi} L_{\rm mp}}e^{-r^2/2 L_{\rm mp}^2}\sqrt{\frac{2}{h_{\rm QD}}}\sin\left(\frac{\pi z}{h_{\rm QD}}\right),
\label{eq:FMP}
\end{equation}
where $r^2\!=x^2+y^2$. Here, $L_{\rm mp}$ is the variational parameter, unlike in Sec.\ref{Sec.Results}.  
The classical turning radius of this oscillator is referred to as $R_{\rm cls} = \sqrt{2} \xi_0$. We take the cylindrical surface of radius $R_{\rm cls}$ and height $h_{\rm QD}$ to be the boundary between the QD and the surrounding layer containing Mn.

The ground state variational energy of the MP is obtained by calculating the expectation value of the total Hamiltonian, 
Eq.~\ref{eq:FMP}. For a fixed carrier and Mn spin configuration, we obtain 
\begin{align}
E(L_{\rm mp}) =  \frac{\hbar^2}{2 m^* L_{\rm mp}^2} + \frac{\hbar^2 \pi^2}{2 m_z^*  h_{\rm QD}^2}+
\frac{1}{2} m^* \omega^2 L_{\rm mp}^2 \nonumber\\
 - \frac{ |N_0 \beta| x_{\rm Mn} S J} {3} e^{-R_{\rm cls}^2/L_{\rm mp}^2},
\label{eq:Evar}
\end{align}
where the first term is the kinetic energy in the $x-y$~plane, 
the second term is the kinetic energy along the $z-$axis with effective mass 
$m^*_z = m_0/\left(\gamma_1-2\gamma_2\right)$,
the third term is the confinement energy and the last term is the exchange energy. For details on the derivation of Eq.~\ref{eq:Evar} see Refs.~\onlinecite{Barman2015:PRB,Pientka2015:PRB}. To obtain the optimal width, we numerically 
minimize Eq.~\ref{eq:Evar} with respect to $L_{\rm mp}$ and obtain an approximate 
MP wave function and energy.  
In the limit of no Mn ($x_{\rm Mn} \rightarrow 0$), we recover the non-magnetic width ($L_{\rm mp}\to \xi_0$). 

Using this model, we find $L_{\rm MP}/\xi_0 =1.06,1.21~{\rm and}~1.37$ for $x_{\rm Mn} = 1 \%, 3\%~{\rm and}~5\%$, respectively.  
Thus, the magnetic width increases with increasing $x_{\rm Mn}$.
The heavy hole wave function expands to increase the overlap between the carrier spin and the Mn spins to lower the total energy. The exchange interaction gives rise to a \emph{delocalization} effect.

The delocalization effect can also be studied in QDs with multiple occupancies. 
We model this by considering a 2D QD with harmonic confinement, and occupied by two heavy holes. 
As before, the Mn ions are distributed continuously in the space surrounding the confined region 
($R \geq R_{\rm cls}$). Using the linear variational method, 
we numerically diagonalize the QD Hamiltonian (containing Coulomb interaction) and obtain the heavy hole's energies and wave function. 
We found that with increasing $x_{\rm Mn}$, the electronic density of the ground state delocalizes 
from the QD center to the QD boundary. 
Using the above parameters with a relative permittivity $\varepsilon =9.3$, \cite{Said1990:pssb}
we find that the weight of the singlet
electronic density outside
$R_{\rm cls}$ to be  $\approx 23\%$ for $x_{\rm Mn} = 0\%$ and  $\approx 40\%$ for $x_{\rm Mn} = 3\%$, 
where half of the Mn spins outside of the QD point up on one side of the line that divides the QD in half, 
and down on the other side of this line.
This indicates that it is energetically favorable for the heavy holes to increase their overlap with the magnetic ions outside the QD. 
A comprehensive discussion of our description of multiply occupied QD is in preparation for a separate publication.

\section{Comparison to Experiment}
 Detailed comparison of results of the self-consistent method to experimental findings for CdMnTe QDs embedded in a bulk semiconductor is not straightforward. Magnetic polaron signatures are typically detected in photoluminescence (PL), where the optical transitions are between levels derived from the conduction and valence bands. The transition energy, h$\nu$, depends on QD geometry, which typically has significant uncertainties. 
 One should also take into account the Varshni shift when analyzing the temperature variation of h$\nu$.\cite{Maksimov2000:PRB} Moreover, our calculated $E_{\rm MP}$ values cannot be directly compared to MP signatures seen in those QDs, where recombination time is smaller or comparable to MP formation time.\cite{Klopotowski2013:PRB}

Hence, we can only compare general trends and orders of magnitude of $E_{\rm MP}$. 
For example, Maksimov et al.\cite{Maksimov2000:PRB}, obtain an estimate of 10.5~meV at $T\simeq 2$~K for a Cd$_{0.93}$Mn$_{0.07}$Te QD formed by fluctuation of a quantum-well width. This value corresponds well to the range presented in Fig.~\ref{emp2}, taking into account the different confinements,\cite{noteShape} and the fact that at the nominal $x_{\rm Mn}=7\%$, some Mn ions are not magnetically active.\cite{Furdyna1988:JAP} 

K{\l}opotowski et al.~obtained $E_{\rm MP}$ from PL of individual, self-assembled Cd$_{1-x{\rm Mn}}$Mn$_{x{\rm Mn}}$Te QDs at $T=8$ K.\cite{Klopotowski2011:PRB}  Their values, $E_{\rm MP}= 9.4$ and $13$ meV for $x_{\rm Mn}=3.5\%$ and $20\%$, respectively, are in the range of our results (the latter high nominal $x_{\rm Mn}$ corresponds to a much smaller effective $x_{\rm Mn}$).

Finally, we note that the wavefunction localization has been seen experimentally for DMS quantum wells.\cite{Akimov2017:PRB}

\section{Discussion and Conclusion}
We have presented preliminary results from a robust numerical method to calculate electronic and magnetic
properties of self-assembled DMS quantum dots, in which equilibrium magnetic polarons are formed. 
The method is based on a well-controlled mean-field approach. 
The obtained values of magnetic polaron binding energy in function of temperature are in the correct range.
Our method allows to calculate spatial profiles of Mn-ion magnetization,
as well as localization of wave-function envelopes.

To our knowledge, little effort has been devoted so far to this level of theoretical description 
of MP formation in quantum dots, accounting for  self-consistency.
A similar, but not identical, approach has been presented in Ref.~\onlinecite{Villegas-Lelovsky2011:PRB}.
Results from the two methods seem to differ at low Mn concentrations, where the approach of Ref.~\onlinecite{Villegas-Lelovsky2011:PRB}
does not predict lifting of ground state degeneracy. A detailed comparison requires further 
calculations. A self-consistent mean-field model adapted to spherical Quantum Dots has been presented in Ref.\onlinecite{Bhattacharjee1997:PRB}.

As a next step of development of the numerical method, we will
consider ``delocalization" of envelopes. 
Here, we have presented a simulation of this effect by a simple model for singly-occupied QDs at zero temperature.
We have also reported briefly on the prediction of a related effect in double-occupied QD. This result has been 
obtained with a variational method taking into account Coulomb interaction between confined carriers.
The variational method will be discussed in a separate publication.

\begin{acknowledgments}
D.R., A.P., and R.O.~acknowledge financial support of U.S.~DoE, grant DE-SC00004890.
\end{acknowledgments}

\end{document}